\documentclass[10pt]{article}
\usepackage[margin=1in]{geometry}
\usepackage{amsmath}
\usepackage{abstract}
\usepackage{amsfonts,color}
\usepackage{amssymb}
\usepackage{graphicx}
\usepackage{hyperref}
\begin{document}
\title{\bf{Analogue tachyon in Jeans Cloud}}
\author
{Satadal Datta$^{1,2}$\\
$^1$Harish-Chandra research Institute, Chhatnag Road, Jhunsi, Allahabad-211019, INDIA\\
$^2$Homi Bhabha National Institute, Training School Complex,\\
Anushaktinagar, Mumbai - 400094, India\\
\date{}
Email: satadaldatta1@gmail.com}
\maketitle
\begin{abstract}
We study the linear perturbations in a stable Jeans cloud, i.e; the dimension of the cloud is less than the Jeans length. We find that the linear perturbation of density in such a system obeys a wave equation in acoustic analogue of Minkowski space-time which is similar to Klein-Gordon equation for tachyon field in Minkowski space-time, i.e; Klein-Gordon equation with negative mass-squared term in a flat space time background. We further find the analogy with tachyon field for linear perturbation of density by studying linear perturbations in a stable cloud made of Bose-Einstein condensate as dark matter.
\end{abstract}
\section{Introduction}
Tachyons are hypothetical particles which move faster than light, proposed in the last century\cite{a}\cite{b}. Tachyons violate causality\cite{b1}\cite{b2} and that leads to paradox\cite{c}. In the case of tachyon exchange between two observers, one of them is moving with uniform velocity with respect to the other one;  transcendent tachyons\cite{d} emitted from one reference frame 'time-travel' in the past of the other, leading to the possibility of causal loops and thus leading to logical paradox. In the description of field theory, tachyon field can be thought of as a field having negative mass-squared term in the Lagrangian\cite{d}\cite{e}, i.e; faster than light particle have imaginary rest mass but energy and momentum are always real.\\ 
Unruh's pioneering work\cite{f} opened a new field, named analogue gravity in theoretical physics as well as in experimental physics which is about producing black hole event horizon like behaviour in lab. The linear perturbation of velocity potential in a irrotational and inviscid fluid, where steady state solutions exist in the background, obeys mass-less scalar field equation in curved space-time background, i.e; sound mimics light propagation in curved space time background. If the background steady state velocity of the fluid crosses the sound sped somewhere, the nearby region have an astounding similarity with the black hole or white hole horizon depending on the direction of the flow\cite{g}\cite{h}. Works are done in relativistic framework as well\cite{i}\cite{j}. Analogue gravity also has immense astrophysical significance, in different astrophysical accretion system linear perturbation of constant of motions, i.e; mass accretion rate\cite{k}-\cite{n}and Bernoulli's constant\cite{n}-\cite{p} over a steady state background solution also mimic mass-less scalar field propagation in curved space-time background and also this methodology is useful to study the stability of such systems; sonic horizon is formed in the case of transonic accretion. \\
Linear perturbation of different suitable quantities in different astrophysical scenarios give rise to phenomenon of analogue gravity. Here we consider Jeans cloud\cite{q} as our astrophysical system and we study the behaviour of linear perturbation of density in such a system. Linear perturbation of density mimic tachyon field propagation in flat-space time, i.e; linear perturbation of density has speed faster than sound. The acoustic space-time is flat, i.e; Minkowski like because the comoving frame is same at every point in Jeans cloud and additionally Jeans cloud is assumed to be of uniform density and uniform temperature which makes it a perfect candidate in producing acoustic analogue space time for Minkowski space-time.\\
In recent years, Bose-Einstein condensate in different potential configurations have drawn a great practical interest in the community because repulsively interacting BEC under T-F approximation gives equations similar to inviscid irrotational fluid equations\cite{r}. There are already some existing works on analogue gravity in Bose-Einstein Condensates\cite{s}\cite{t}. BEC systems are now thought to produce analogue black holes in laboratory. BEC has astrophysical significances too. Gravity is introduced as an interaction in BEC to describe a class of self-gravitating objects\cite{u}-\cite{w}.  There are also some cold star models where the constituent of a star is BEC and Newtonian gravity is an interaction among the particles of the condensate\cite{x}-\cite{z}. Bose-Einstein condensate is also thought of as a candidate of dark matter. Dark Matter made of bosons of mass of the order of 1eV or less has critical condensation temperature which is greater than the temperature of the universe at all times and hence giving rise to Bose-Einstein condensation in the early universe\cite{z1}. To explain small scale structures by cold dark matter model (CDM), BEC is thought of as possible dark matter candidate\cite{y1}. Non-relativistic Gross-Pitaevskii equation coupled to the Poisson's equation in Newtonian gravity is modeled as the governing equation for BEC as a dark matter candidate\cite{x1}\cite{w1}.\\
We consider a self-gravitating cloud consisting of BEC, we study the stability of such a cloud by introducing linear perturbations in density and velocity. The linear perturbation of density of condensate obeys equation of motion with negative mass squared term. Hence the analogy with the tachyon is drawn as conclusion.\\
We discuss the invariance of acoustic metric which is analogous to Minkowskii metric and introduce a hypothetical coordinate transformation under which the analogue acoustic metric is invariant and then we discuss the violation of causality by the analogue tachyon field. 
\section{Linear Perturbations in Jeans Cloud} 
Let's consider a self-gravitating Jeans cloud, the governing equations are in general as follows
\begin{equation}
\partial_t\rho+\vec{\nabla}.(\rho\vec{v})=0
\end{equation}
\begin{equation}
\partial_t\vec{v}+\vec{v}.\vec{\nabla}\vec{v}=-\frac{\vec{\nabla}p}{\rho}-\vec{\nabla}\Psi
\end{equation}
\begin{equation}
\nabla^2\Psi=4\pi G\rho
\end{equation}
where $p,~\rho,~\vec{v}$ are fluid pressure, density and velocity respectively; $G$ is universal Gravitational constant and $\psi$ is gravitational potential.
Let 
\begin{align*}
&p=p_0+p'(\vec{x}, t)\\
&\rho=\rho_0+\rho'(\vec{x}, t)\\
&\vec{v}=\vec{v}'(\vec{x}, t)\\
&\Psi=\Psi_0+\Psi'(\vec{x}, t)
\end{align*}
where $p_0,~\rho_0,~\Psi_0$ are uniform pressure, uniform density and uniform gravitational potential of the medium respectively, namely Jeans Swindle\cite{v1}. The background medium is static. $p',~\rho',~\vec{v'}$ and $\Psi'$ are the linear perturbations in pressure, density, velocity and gravitational potential respectively. Euation (1)-equation (3) respectively takes form:
\begin{equation}
\partial_t\rho'+\vec{\nabla}.(\rho_0\vec{v}')=0
\end{equation}
\begin{equation}
\partial_t\vec{v}'+\frac{c_{s0}^2}{\rho_0}\vec{\nabla}\rho'+\vec{\nabla}\Psi'=0
\end{equation}
\begin{equation}
\nabla^2\Psi'=4\pi G\rho'
\end{equation}
where $c_{s0}$ is the isothermal sound speed\cite{v1}, given by 
\begin{align*}
c_{s0}^2=\frac{k_BN_AT}{M}
\end{align*}
$k_B,~N_A,~T$ and $M$ are the Boltzmann constant, Avogadro Number, uniform temperature of the medium and molar weight of the material of the cloud. Taking a second partial derivative in equation (4) and using equation (5) and equation (6) we find
\begin{equation}
(\square_A+\mu^2c_{s0}^2)\rho'=0
\end{equation}
where $\square_A$ is the acoustic analogue of d'Alembertian operator in which light speed is replaced by sound speed, hence
\begin{align*}
\square_A=-\frac{1}{c_{s0}^2}\frac{\partial^2}{\partial t^2}+\nabla^2
\end{align*}
and $\mu^2$ is $\frac{4\pi G\rho_0}{c_{s0}^4}$.\\
The analogue acoustic metric in general, is given by\cite{h}
\begin{equation}
ds^2=\frac{\rho}{c_{s0}}\left[-c_{s0}^2dt^2+(dx^i-v_0^idt)\delta_{ij}(dx^j-v_0^jdt)\right]
\end{equation}
This metric reduces to the acoustic analogue of Minkowskii metric when the background speed of the medium, $v_0$ is zero and the medium is homogeneous and isothermal. Therefore equation (7) represents Klein-Gordon equation\footnote{Here we have taken  $\hbar\rightarrow1$}\cite{u1} with a negative mass squared term in the acoustic analogue of Minkowskii space-time. For completeness, one can write the corresponding action for linear perturbation of density as 
\begin{equation}
S=\frac{1}{2}\int d^4x \left(\dot{\rho'}(x)\dot{\rho'^*}(x)-c_{s0}^2\vec{\nabla }\rho'(x).\vec{\nabla} \rho'^*(x)+\mu^2c_{s0}^4\rho'(x)\rho'^*(x)\right)
\end{equation}
One can construct Hamiltonian also,
\begin{equation}
H=\frac{1}{2}\int d^3\vec{x}\left(\dot{\rho'}(x)\dot{\rho'^*}(x)+c_{s0}^2\vec{\nabla }\rho'(x).\vec{\nabla }\rho'^*(x)-\mu^2c_{s0}^4\rho'(x)\rho'^*(x)\right)
\end{equation}
The negative mass squared term in the Hamiltonian implies the instability of analogue tachyon vacuum, hence analogue Tavhyon vacuum isn't possible. We know that Jeans cloud is stable under linear perturbation having wave number, $k>k_J$ and $k_J=\sqrt{\frac{4\pi G\rho_0}{c_{s0}^2}}$\cite{v1}. Therefore, $\mu^2=\frac{k_J^2}{c_{s0}^2}=\frac{4\pi^2}{\lambda_{ms}^2c_{s0}^2}$. $\lambda_{ms}$ is Jeans length. Taking $\rho'=\tilde{\rho}e^{i(\vec{k}.\vec{x}-\omega t)}$ where $\tilde{\rho}$ is the amplitude of the density perturbation, using equation (7), we also get
\begin{equation}
\omega^2=c_{s0}^2(k^2-k_j^2)=c_{s0}^2k^2-\mu^2c_{s0}^4
\end{equation}
For Jeans cloud, stable under linear perturbations, $k>k_J$, we get the same known conclusion\cite{v1}. Hence the energy and momentum of analogue tachyon are real but the rest mass is purely imaginary in stable Jeans cloud. In the same conventional fashion we can also work in a unit just like Natural unit by taking $\hbar\rightarrow1,~c_{s0}\rightarrow1$ if we wish. Hence
\begin{equation}
\omega(\vec{k})^2=\vec{k}^2-\mu^2
\end{equation}
\section{Cloud Made of BEC}
Now, we consider interacting BEC as the constituent of a non-rotating cloud, the system can be thought of as a cold $(T=0K)$ dark matter cloud. The cloud is self-gravitating. For an interacting BEC famous Gross-Pitaevskii equation reads as 
\begin{equation}
i\hbar\frac{\partial \Phi(\vec{x},t)}{\partial t}=(-\frac{\hbar^2}{2m}\nabla^2+V_{ext}(\vec{x})+g|\Phi(\vec{x},t)|^4)\Phi(\vec{x},t)
\end{equation} 
where $\Phi(\vec{x},t)$ is the wave function of the condensate, $V_{ext}$ is the external potential, $g$ is two body interaction coefficient related to s-wave scattering cross-section.
\begin{align*}
g=\frac{4\pi\hbar^2a}{m}
\end{align*}
where $a$ is the scattering length. $g$ is positive for repulsive interaction and negative for attractive interaction. We assume repulsive interaction.\\
Now we consider self-gravity. There is no external potential, the only potential, arising from the gravitational interaction among the particles, is Newtonian gravity satisfying Poisson's equation\cite{z}.
\begin{equation}
\nabla^2\Psi=4\pi G\rho
\end{equation}
and density
\begin{align*}
\rho(\vec{x},t)=mn(\vec{x},t)=m|\Phi(\vec{x},t)|^2
\end{align*} 
where $n(\vec{x},t)$ is the number density of the bosons.
Hence Gross-Pitaevskii equation can be rewritten as
\begin{equation}
i\hbar\frac{\partial \Phi(\vec{x},t)}{\partial t}=(-\frac{\hbar^2}{2m}\nabla^2+\Psi(\vec{x})+g|\Phi(\vec{x},t)|^4)\Phi(\vec{x},t)
\end{equation}
Now writing $\Phi(\vec{x},t)=\sqrt{n(\vec{x},t)}e^{iS(\vec{x},t)}$ and $\vec{v}(\vec{x},t)=\frac{\hbar}{m}\vec{\nabla}S(\vec{x},t)$\cite{r}, we get continuity equation and irrotationality condition
\begin{equation}
\partial_t\rho+\vec{\nabla}.(\rho\vec{v})=0
\end{equation}
\begin{equation}
\vec{\nabla}\times\vec{v}=0
\end{equation}
Now using equation (15) and using T-F approximation for repulsive interaction, we get Euler equation
\begin{equation}
\frac{\partial\vec{v}}{\partial t}+\vec{v}.\vec{\nabla}\vec{v}=-\vec{\nabla}\Psi-\frac{\vec{\nabla}p}{\rho}
\end{equation}
where $p=\frac{1}{2}gn^2=\frac{g}{2m^2}\rho^2$. Thus we get all the fluid equations. Now in the same way as in the previous section we  introduce linear perturbations in density, velocity and gravitational potential (background uniform gravitational potential$=\Psi_0$) of the fluid over a uniformly dense (background uniform density$=\rho_0$), static and isothermal background BEC medium, i.e; using Jeans swindle. Next after similar calculations, we get Klein-Gordon equation for linear perturbation of density with negative mass squared term. Hence we find analogue tachyon field. The mass squared term
\begin{align*}
\mu^2=\frac{4\pi G\rho_0}{c_{s0}^4}
\end{align*}
where 
\begin{equation}
c_{s0}^2=\frac{gn}{m}=\frac{g\rho_0}{m^2}
\end{equation}  
The cloud is stable under linear perturbation of wave number $k$ such that $k>k_J$. \\
Where
\begin{equation}
k_J^2=\frac{4\pi G\rho_0}{c_{s0}^2}=\mu^2c_{s0}^2=\frac{Gm^3}{\hbar^2a}
\end{equation}
Hence $k_J$ is independent of background density of the medium.
\section{Analogue Tachyon Particle}
To explore the particle aspect of analogue tachyon field, we have to quantize the field. We treat the linear perturbation of density, $\rho'$ as an operator and introduce creation and annihilation operator as below
\begin{align*}
&[\hat{a}(\vec{k}, t),a(\vec{k'}, t)]=0\\
&[\hat{a}^\dagger(\vec{k}, t),a^\dagger(\vec{k'},t)]=0\\
&[\hat{a}(\vec{k}, t),a^\dagger(\vec{k'}, t)]=\delta^3(\vec{k}-\vec{k'})
\end{align*}
We write
\begin{equation}
\rho'(\vec{x},t)=\int_{|\vec{k}|>k_J,\omega>0}\frac{d^3\vec{k}}{(2\pi)^{\frac{3}{2}}(2\omega(\vec{k}))^{\frac{1}{2}}}[\hat{a}(\vec{k})e^{i(\vec{k}.\vec{x}-\omega t)}+\hat{a}^\dagger(\vec{k})e^{-i(\vec{k}.\vec{x}-\omega t)}]
\end{equation}
$\rho'(\vec{x},t)$ satisfies equation (7).\\
No particle vacuum state, $|0\rangle$ is defined by
\begin{equation}
\hat{a}(\vec{k},t)|0\rangle=0
\end{equation}
From equation (7), the field equation corresponds to 2nd order time derivative, therefore from equation (7),we can say that if at $t=0$ the analogue tachyon field is real then it would be real field at any time t. Therefore $\rho'(\vec{x},t)$ is Hermitian.
Hence
\begin{align*}
\hat{a}^\dagger(\vec{k})=\hat{a}(-\vec{k})
\end{align*}
Omitting infinite term related to zero point oscillation, we get after some calculations (not shown for brevity)
\begin{equation}
H=\int_{|\vec{k}|>k_J,\omega>0}d^3\vec{k}~\omega(\vec{k})~\hat{a}^\dagger(\vec{k})\hat{a}(\vec{k})
\end{equation}
As an immediate consequence, we get some results as below
\begin{align*}
&H|0\rangle=0\\
&H\hat{a}^\dagger(\vec{k},t)|0\rangle=\omega(\vec{k})\hat{a}^\dagger(\vec{k},t)|0\rangle
\end{align*}
$n$ particle state with momentum $\vec{k_1},\vec{k_2},....\vec{k_n}$ can be written as 
\begin{equation}
|\vec{k_1},\vec{k_2},....\vec{k_n}\rangle=\hat{a}^\dagger(\vec{k_1},t)\hat{a}^\dagger(\vec{k_2},t)...\hat{a}^\dagger(\vec{k_n},t)|0\rangle
\end{equation}
and 
\begin{equation}
H|\vec{k_1},\vec{k_2},....\vec{k_n}\rangle=(\omega(\vec{k_1})+\omega(\vec{k_2})+...+\omega(\vec{k_n})|\vec{k_1},\vec{k_2},....\vec{k_n}\rangle
\end{equation}
\section{Analogue Tachyon Kinematics}
Equation (11) gives a dispersion relation. The energy of analogue tachyon is real, i.e; $\omega^2>0$. Therefore
\begin{equation}
\frac{\mu^2c_{s0}^2}{k^2}<1
\end{equation}
The group velocity, $v_{g}$\footnote{$|\vec{v_g}|=v_g,~|\vec{k}|=k$} of such a density wave is given by
\begin{equation}
v_g=\frac{d\omega}{dk}=\frac{c_{s0}}{\sqrt{1-\frac{\mu^2c_{s0}^2}{k^2}}}
\end{equation}
$\Rightarrow$ $v_g>c_{s0}$. Hence negative mass squared density perturbation wave has speed faster than sound. Sound in a medium behaves like light\cite{t1}, hence linear perturbation of density is faster than the analogue of light. Therefore we call it by the name analogue tachyon. \\
Using equation (27), we find
\begin{equation}
k=\frac{\mu v_g}{\sqrt{\frac{v_g^2}{c_{s0}^2}-1}}
\end{equation}
The above expression represents the momentum of analogue tachyon where $\hbar\rightarrow1$. Again expression of energy is given by
\begin{equation}
\omega=\frac{\mu c_{s0}^2}{\sqrt{\frac{v_g^2}{c_{s0}^2}-1}}
\end{equation}
The above expressions are exactly similar with the expression of tachyon energy and momentum.
Energy and momentum are real but the rest mass of analogue tachyon particle is purely imaginary, i.e; $m_0=i\mu$. Energy and momentum increases indefinitely as $v_g$ approaches $c_{s0}$ and as $v_g$ tends to infinity\footnote{$v_g$ can not approach infinity because of universal speed limit, the upper bound of $v_g$ is $c$. In reality as $v_g<<c$, we are in non-relativistic limit} energy of analogue tachyon diminishes and momentum, $k$ approaches it's minimum value $k_J$.\\
\begin{figure}
\center
\includegraphics[scale=0.5,angle=-90]{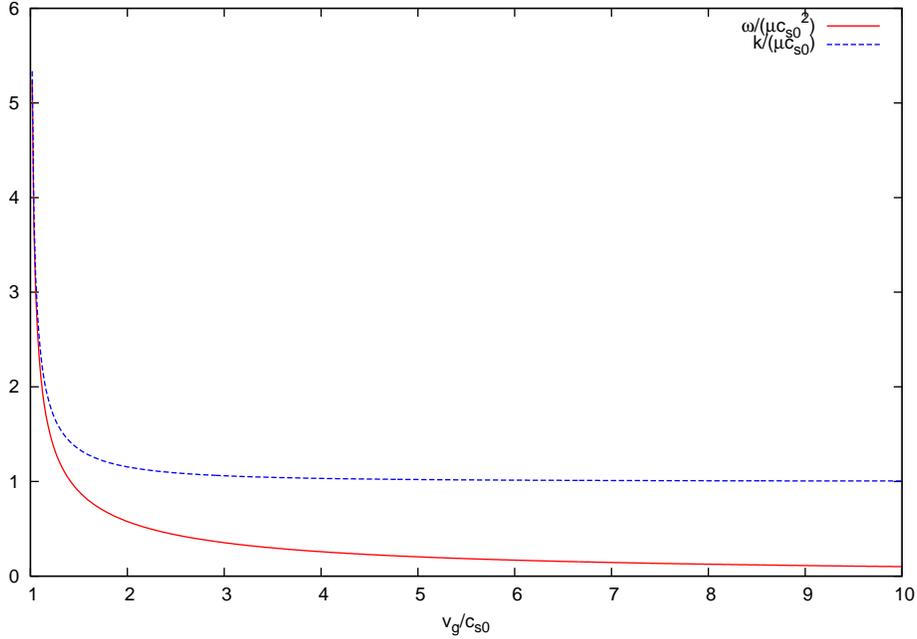} 
\caption{Energy and momentum of analogue tachyon decreases as group velocity approaches infinity.}
\end{figure}
Strictly speaking, the speed of analogue tachyon is bounded below as well as bounded above
\begin{align*}
c_{s0}<v_g<c
\end{align*}
Hence the minimum momentum, $k_{min}$ isn't $k_J$ and minimum energy, $\omega_{min}$ of analogue tachyon is not exactly zero.
\begin{align*}
&k_{min}=\frac{\mu c}{\sqrt{\frac{c^2}{c_{s0}^2}-1}}\\
&\omega_{min}=\frac{\mu c_{s0}^2}{\sqrt{\frac{c^2}{c_{s0}^2}-1}}
\end{align*}
In the non-relativistic limit, i.e; for $c>>c_{s0}$, $k_{min}\rightarrow k_J$ and $\omega_{min}\rightarrow 0$.
\section{Notion of Reference Frame}
The acoustic analogue of Minkowski metric is given by
\begin{equation}
ds^2=(\eta_A)_{\alpha\beta}dx^\alpha dx^\beta=-c_{s0}^2dt^2+d\vec{x}^2
\end{equation}
Index $\alpha,~\beta$ runs from $0$ to $3$ representing the time coordinate and three space coordinates respectively, i.e; $x^\alpha=[c_{s0}t,~x,~y,~z]$.
$c_{s0}$ is the uniform sound speed in the isothermal medium. $(\eta_A)_{\alpha\beta}$ is the acoustic analogue of Minkowski metric and it looks exactly like the Minkowski metric.\\
An analogue tachyon is space-like when we put $d\vec{x}=\vec{v_g}dt$ and $v_g>c_{s0}$ and also similarly 4-momentum of analogue tachyon particle\footnote{If we introduce 4-momentum concept similar to special relativity, just by replacing light speed by sound speed} is space-like which follows from equation (12). Whatever we have done so far is with respect to the comoving frame of the cloud. There is only one comoving frame in this system  because the medium is static (the background speed is zero) everywhere, there is no relative velocity among different parts of the medium. Let's consider a reference frame moving with uniform speed $v_0$, the cloud has uniform velocity $\vec{v_0}$ with respect to the frame. Now we introduce linear perturbations in the medium. The linear perturbations in density, pressure and gravitational potential looks same as before except the velocity one.
\begin{equation}
\vec{v}=\vec{v_0}+\vec{v'}(\vec{x'},t')
\end{equation} 
We consider the comoving frame to be unprimed frame as done before and the coordinates in this frame which is moving with velocity $-\vec{v_0}$ with respect to the comoving frame are primed. Basically, this two frames are related to each other via a Galiean transformation, hence $t=t'$. The fluid equations in terms of the perturbed quantities read as 
\begin{equation}
\partial_t'\rho'+\vec{\nabla'}.(\rho_0\vec{v}'+\vec{v_0}\rho')=0
\end{equation}
\begin{equation}
\partial_t'\vec{v}'+\vec{\nabla'}(\vec{v_0}.\vec{v'})+\frac{c_{s0}^2}{\rho_0}\vec{\nabla'}\rho'+\vec{\nabla'}\Psi'=0
\end{equation}
\begin{equation}
\nabla'^2\Psi'=4\pi G\rho'
\end{equation}
Taking a second partial time derivative in equation (32), using equation (33) and equation (34)
\begin{equation}
\partial_t'^2(\rho')-c_{s0}^2\nabla'^2\rho'-4\pi G\rho_0\rho'+\vec{\nabla'}.(\vec{v_0}\partial_t')\rho'-\vec{\nabla'}.(\rho_0\vec{v_0}.\vec{\nabla'})\vec{v'}=0
\end{equation}
The last term in the LHS can be written as
\begin{align*}
&\partial'_j(\rho_0v_0^i\partial'_i)v'^j\\
&\because {\rm\rho_0,~v_0^i~are~constants}\\
&=v_0^i\partial'_i\partial'_j(\rho_0v'^j)\\
&=-\partial'_iv_0^i\partial_t'(\rho')-\partial'_iv_0^iv_0^j\partial'_j\rho'
\end{align*}
Equation (35) reduces to
\begin{equation}
(\partial'_\alpha~f^{\alpha\beta}\partial'_\beta+\mu^2c_{s0}^2)\rho'=0
\end{equation}
where 
\begin{equation}
f^{\alpha\beta}=\frac{1}{c_{s0}^{2}}\begin{bmatrix}
-1 & \vdots & -v_{0}^{j} \\
\cdots&\cdots&\cdots\cdots \\
-v_{0}^{j}&\vdots & c_{s0}^{2}\delta^{ij}-v_{0}^{i}v_{0}^{j}
\end{bmatrix}
\end{equation}
and
\begin{align*}
\mu^2=\frac{4\pi G\rho_0}{c_{s0}^4}
\end{align*}
Now, we can find the corresponding acoustic metric by using\cite{g}
\begin{align*}
f^{\alpha\beta}=\sqrt{-g}g^{\alpha\beta}
\end{align*}
We find
\begin{equation}
g_{\alpha\beta}=\frac{1}{c_{s0}}\begin{bmatrix}
 -(c_{s0}^{2}-v_{0}^{2}) & \vdots & -v_{0}^{i} \\
\cdots&\cdots&\cdots\cdots \\
-v_{0}^{i}&\vdots &\delta_{ij}
\end{bmatrix}
\end{equation}
Corresponding acoustic metric is given by
\begin{equation}
ds^{2}=g_{\alpha\beta}dx^\alpha dx^\beta=\frac{1}{c_{s0}}\left[-(c_{s0}^{2}-v_{0}^{2})dt'^{2}-2dt'\vec{v_{0}}.d\vec{x'}+d\vec{x'}^{2} \right]
\end{equation}
Putting $v_0$ to be $0$, the above metric is identical with the metric in equation (30) except for the conformal factor. From equation (30), this metric can also be derived by using Galilean coordinate transformation\cite{s1}\cite{r1} between primed and unprimed coordinate systems.
\begin{equation}
\vec{x}=\vec{x'}-\vec{v_0}t'
\end{equation}
The above metric without the conformal factor in front of it can be realized from geometrical acoustic approximation, i.e; considering the sound cones dragged by the fluid motion\cite{h}. Using the Galilean coordinate transformation, i.e; equation (40) in equation (7), equation (36) can also be derived. There we have to treat fluid density, gravitational potential as scalars under Galilean transformation. The sound speed measured from the primed frame follows the Galilean velocity addition rule, i.e; the sound speed is added along $\vec{v_0}$ and subtracted along $-\vec{v_0}$. This conclusion can also be drawn by putting $ds^2$ to be zero. From equation (38), the signature of this metric is $(-,+,+,+)$ until and unless the speed of the primed frame with respect to the unprimed one is less than the sound speed. If $v_0$ is greater than the sound speed the signature of the metric changes. Hence the nature of the acoustic metric is very much dependent on the reference frame which are obtained via Galilean coordinate transformation.\\
Now let's introduce a coordinate transformation under which the acoustic metric is invariant\cite{q1}.
\begin{equation}
X^\alpha=(L_A)^\alpha~_\beta x^\beta+a^\alpha
\end{equation}
$a^\alpha$ represents constant translation along space-time coordinates. 
$(L_A)^\alpha~_\beta$ satisfies the following property due to invariance of acoustic metric under such transformation.
\begin{equation}
(L_A)^\alpha~_\gamma(L_A)^\beta~_\delta(\eta_A)_{\alpha\beta}=(\eta_A)_{\gamma\delta}
\end{equation}
$(L_A)^\alpha~_\beta$ represents two disjoint set of transformations. One is rotation
\begin{equation}
(L_A)^i~_j=R_{ij},~(L_A)^i~_0=(L_A)^0~_i=0,~(L_A)^0~_0=1
\end{equation}
where index $i,~j$ run over three spatial coordinates. $R_{ij}$ is orthogonal rotation matrix having determent to be unity.
Another set of coordinate transformation is boost which is like Lorentz boost except the light speed is replaced by sound speed.
\begin{equation}
(L_A)^i~_j=\delta_{ij}+V_iV_j\frac{(\Gamma_A-1)}{\vec{V}^2}
\end{equation}
\begin{equation}
(L_A)^i~_0=(L_A)^0~_i=\Gamma_AV_i
\end{equation}
where $\Gamma_A$ is given by
\begin{align*}
\Gamma_A=\left(1-\frac{\vec{V}^2}{c_{s0}^2}\right)^{-\frac{1}{2}}
\end{align*}
$\vec{V}$ is a vector having norm less than sound speed. We impose the new set of coordinates to be represented by real numbers. The above transformations exactly look like Lorentz transformations and hence form group, i.e; similar to inhomogeneous and homogeneous Lorentz group. We may call it acoustic analogue of Lorentz transformation.\\
One important point to mention is that this is not a coordinate transformation of actual space-time. This coordinate transformation; we introduce for mathematical completeness just to keep the acoustic analogue of Minkowski metric invariant under coordinate transformation. The acoustic metric isn't invariant under actual space-time coordinate transformation, i.e; Lorentz transformation. We are basically imposing the invariance of the acoustic analogue of Minkowski metric. We imagine this hypothetical coordinate transformation and as a consequence we find some interesting results discussed in the next section. Again the analogue of Lorentz transformation depend on $c_{s0}$, i.e; the temperature of the cloud.
\section{Violation of Causality}
Let's consider the reference frame comoving with the cloud and a hypothetical reference frame connected via the following coordinate transformation.
\begin{align*}
&X^0=\Gamma_A(x^0-\frac{V_1}{c_{s0}}x^1)\\
&X^1=\Gamma_A(x^1-\frac{V_1}{c_{s0}}x^0)\\
&X^2=x^2\\
&X^3=x^3
\end{align*}
This transformation is a special case of analogue Lorentz boost where $\vec{V}=(V_1,0,0)$. This relation can be obtained from equation (44) and equation (45). Hence here $\Gamma_A=\left(1-\frac{{V_1}^2}{c_{s0}^2}\right)^{-\frac{1}{2}}$. Now let's consider two points $P$ and $Q$ along $x-$ axis in the medium. Ripple of density perturbation generates at point $P$ and is detected at point $Q$. The time difference between two events in the comoving frame is
\begin{equation}
x^0_Q-x^0_P=\delta x^0=c_{s0}\frac{(x^1_Q-x^1_P)}{v_g}
\end{equation}
where $x^1_Q>x^1_P$.\\
In the hypothetical frame, the time gap between these two events is given by 
\begin{align}
&\delta X^0=X^0_Q-X^0_P=\Gamma_A(x^0_Q-\frac{V_1}{c_{s0}}x^1_Q)-\Gamma_A(x^0_P-\frac{V_1}{c_{s0}}x^1_P)\\
&=\Gamma_A(1-\frac{V_1v_g}{c_{s0}^2})\delta x^0
\end{align}
$\because~v_g>c_{s0}$, we can choose $V_1$ and $v_g$ in such a way that $\frac{V_1v_g}{c_{s0}^2}>1$ (analogue of transcendent tachyon), i.e; as a consequence,  $\delta X^0$ becomes negative. Hence in the hypothetical reference frame, the causal ordering between these two events is reversed. Basically, the analogue tachyon time-travel in the past in the hypothetical reference frame. This results is similar to the Tolman paradox\cite{p1}.\\
Under this hypothetical coordinate transformation the energy and momentum also transform exactly like tachyon particle\cite{d} except the light speed is replaced by sound speed everywhere. In the same fashion to explore the particle aspect\cite{o1}of analogue tachyon, one can write equation (21) in analogue Lorentz transformation invariant form\cite{d}.
\section{Summary and Conclusions}
We have shown that linear perturbation of density in a self-gravitating cloud of uniform density and uniform temperature propagates like tachyon in the acoustic analogue of Minkowski space-time, i.e; the speed of such density wave is faster than the sound speed in the medium. The negative mass squared term appearing in the equation of motion of the linear perturbation of density is the manifestation of self-gravity. Self-gravity term is actually responsible for such a dispersion relation that gives group velocity of density wave to be greater than the sound speed. We have also explored particle nature of analogue tachyon.\\
Incorporating gravity, hence finding analogue tachyon in BEC system; has also experimental value. An experimental set up is proposed\cite{n1}; where a standing wave cavity of infra-red laser formed by gold Casimir shield traps BEC atoms. The alternating dense regions of gold and silver serve as periodic source mass. Thus gravity in sub-millimeter range may be realised. Thus there is also a hope to analyze the behaviour of tachyon particle practically by studying the acoustic analogue of tachyon in laboratory. 

\end{document}